\documentclass[
 reprint,
 superscriptaddress,
 amsmath,amssymb,
 aps,
]{revtex4-2}

\usepackage{graphicx}
\usepackage{amsmath}
\usepackage{mathtools}
\usepackage{amssymb}

\DeclarePairedDelimiter\abs{\lvert}{\rvert}%
\DeclarePairedDelimiter\norm{\lVert}{\rVert}%

\makeatletter
\let\oldabs\abs
\def\abs{\@ifstar{\oldabs}{\oldabs*}}
\let\oldnorm\norm
\def\norm{\@ifstar{\oldnorm}{\oldnorm*}}
\makeatother

\begin{document}

\title{Krylov Complexity and Mixed-State Phase Transition}
\date{\today}

\author{Hung-Hsuan Teh}
\thanks{These authors equally contributed to this work}
\affiliation{The Institute for Solid State Physics, The University of Tokyo, Kashiwa,
Chiba, 277-8581, Japan}
\affiliation{Graduate School of Informatics, Nagoya University, Nagoya, 464-0814, Japan}

\author{Takahiro Orito}
\thanks{These authors equally contributed to this work}
\affiliation{\mbox{Department of Physics, College of Humanities and Sciences, Nihon University, Sakurajosui, Setagaya, Tokyo 156-8550, Japan}}

\begin{abstract}
We establish a unified framework connecting decoherence and quantum complexity. By vectorizing the density matrix into a pure state in a double Hilbert space, a decoherence process is mapped to an imaginary-time evolution. Expanding this evolution in the Krylov space, we find that the $n$-th Krylov basis corresponds to an $n$-error state generated by the decoherence, providing a natural bridge between error proliferation and complexity growth. Using two dephasing quantum channels as concrete examples, we show that the Krylov complexity remains nonsingular for strong-to-weak spontaneous symmetry-breaking (SWSSB) crossovers, while it exhibits a singular area-to-volume-law transition for genuine SWSSB phase transitions, intrinsic to mixed states.
\end{abstract}

\maketitle

\textit{Introduction---}Unavoidable couplings to the environment drive a pure state into a mixed state, often leading to featureless forms such as those at infinite-temperature.
This process, known as decoherence or quantum noise~\cite{gardiner2000,Zurek2003decoherence}, has become a central obstacle to reliable quantum computation~\cite{dennis2002,preskill2018}. 
Robust quantum memories, precisely controllable quantum systems, and error-correction techniques are therefore indispensable for quantum technologies, and their feasibility has recently been demonstrated experimentally~\cite{exp1,exp2}.

Although usually regarded as detrimental, decoherence constrained by symmetry can generate nontrivial structures in mixed states and even mixed-state phase transitions~\cite{coherent-info-1,decohere-SPT,ASPT1,MixedTO1,MixedTO2,MixedTO3,Mixed-state-Hsieh,PhysRevB.111.115141}.
In such settings, symmetries admit two distinct forms---strong and weak~\cite{Buca_2012,Albert_2014,groot2022}---giving rise to novel symmetry-breaking patterns with no analogue in pure-state physics. A prominent intrinsic example is strong-to-weak spontaneous symmetry breaking (SWSSB)~\cite{Ex1,Ex2,Ex3,Ex4,Ex5,Ex6,symmetry-in-OQS,hardness&exp-SWSSB-2}.

Several correlators have been proposed to detect SWSSB~\cite{SWSSB1,SWSSB-fidelity-corr}, including Rényi-1~\cite{Renyi-1-1}, Rényi-2~\cite{SWSSB1,SWSSB-oshikawa}, fidelity~\cite{SWSSB-fidelity-corr}, and Wightman correlators~\cite{Renyi-1-2,Wightman2}.
Some of these satisfy stability theorems~\cite{stability-theorem,SWSSB1,SWSSB-fidelity-corr}, allowing mixed-state phases to be defined and classified.
However, unlike conventional symmetry breaking, e.g. in superconductivity, where order parameters carry clear physical meaning~\cite{Sachdev}, the correlators probing SWSSB are nonlinear in the density matrix $\rho$ and thus difficult to interpret or measure~\cite{hardness&exp-SWSSB-1}.
This motivates a natural question: can SWSSB phase transition be identified directly from the density matrix itself, without reference to specific observables? Moreover, can it be understood as a transition in the \textit{complexity} of $\rho$?

\begin{figure}[t]
\begin{center} 
\includegraphics[width=.45\textwidth]{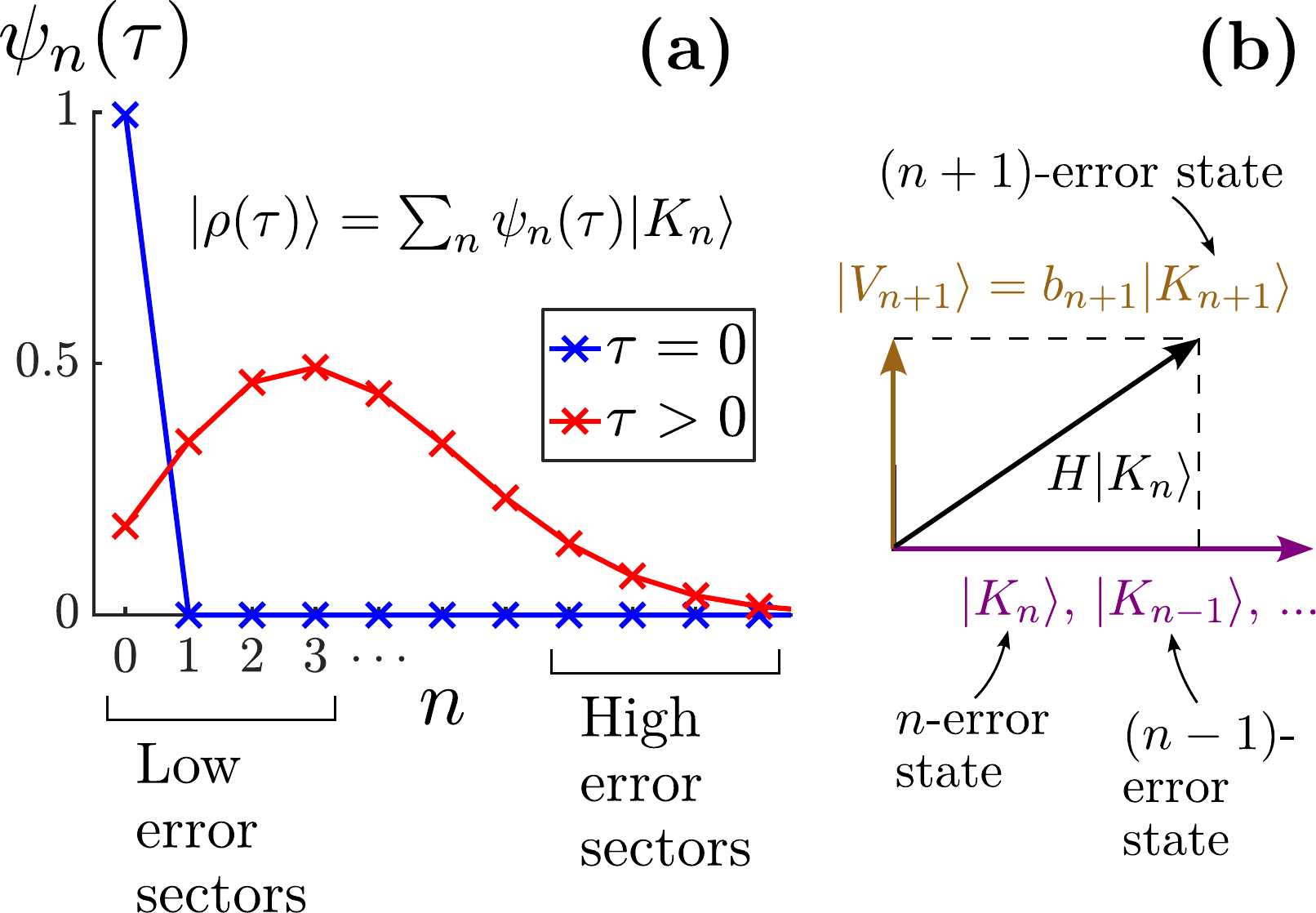}  
\end{center} 
\vspace{-.3cm}
\caption{(a) Schematic illustration of the wave-packet dynamics in the Krylov basis $\vert K_{n}\rangle$, where each $\vert K_{n}\rangle$ represents an $n$-error state generated by decoherence. As time evolves, the wave packet $\psi_{n}(\tau)$ spreads toward higher-error sectors, reflecting the growth of quantum complexity.
(b) Gram–Schmidt construction of the Krylov basis. The $(n+1)$-th basis state $\vert K_{n+1}\rangle$ is obtained by applying the noise operator $H$ to the $n$-error state $\vert K_{n}\rangle$, followed by projecting out all lower-error components $\vert K_{n}\rangle$, $\vert K_{n-1}\rangle$, ...}
\label{Fig1}
\end{figure}

To address these questions,
we outline our central idea here. 
We vectorize the density matrix $\rho$ into a pure state $\vert\rho\rangle$ using the double Hilbert space formalism~\cite{Choi1975,JAMIOLKOWSKI1972}.
Under this mapping, the decoherence channel $\mathcal{E}$ becomes an imaginary-time evolution operator $e^{-H\tau}$, where $H$ is an effective Hamiltonian encoding the noise, and $\tau$ is an imaginary time increasing monotonically with the decoherence strength $p$.
Thus, increasing $p$---where
an SWSSB phase transition may occur---can be viewed as evolving a state in imaginary time. Acting $\mathcal{E}$ on an initial density matrix $\rho_{\text{init}}$, the vectorized decohered state is given by $\vert\mathcal{E}[\rho_{\text{init}}]\rangle \sim e^{-H\tau}\vert\rho_{\text{init}}\rangle \equiv \vert\rho(\tau)\rangle = \sum_{n}(-\tau)^{n}H^{n}\vert\rho_{\text{init}}\rangle/n!$, a superposition of states with different error levels, $\{H^{n}\vert\rho_{\text{init}}\rangle\}$. The subspace spanned by these states is precisely the Krylov space, whose orthonormal basis $\{\vert K_{n}\rangle\}$ provides a natural framework for analyzing how information spreads under decoherence. Expanding as $\vert\rho(\tau)\rangle=\sum_{n}\psi_{n}(\tau)\vert K_{n}\rangle$, the weights $\abs{\psi_{n}}^{2}$ quantify the extent to which the state explores higher-error subspaces (as shown schematically in Fig.~\ref{Fig1} (a)), i.e., its effective complexity.

To quantify this spread more explicitly, we consider  the ``center of mass'' of the wave packet in the Krylov basis, $\mathcal{K}(\tau)=\sum_{n}n\abs{\psi_{n}(\tau)}^{2}$, known as the Krylov complexity. The conventional Krylov complexity $\mathcal{K}(t)$, defined for real-time evolutions, has been widely used as a diagnostic of dynamical chaos~\cite{Krylov-review1,Krylov-review2,Krylov-review3,Krylov-review4}.
In chaotic systems, $\mathcal{K}(t)$ typically exhibits a peak prior to saturation, while in non-chaotic systems such a peak is absent~\cite{BalasubramanianPRD2022,Erdmenger_2023,Alishahiha2025}.
Moreover, the same concept naturally extends to operator dynamics~\cite{Operator-spreading}.
While the Krylov complexity has been firmly established as a sensitive probe of chaos, we ask: can it also detect mixed-state phase transitions?
Here we demonstrate that the answer is positive: SWSSB phase transition manifests as singularities in the Krylov complexity, exhibiting an area-to-volume-law transition.

We refer to the situation where SWSSB occurs only at the boundary of the $p$ domain as a ``crossover’’ rather than a true phase transition.
Numerous decohered systems exhibit such crossovers, including a noisy spin-1/2 1D chain~\cite{SWSSB-oshikawa}, 1D cluster state~\cite{noisy-cluster}, and 2D gauged Hamiltonian~\cite{SWSSB-2Dgauged}, and in the thermodynamic limit the SWSSB phase reduces to a single-point change that is not detectable in practice.
An analogy can be drawn from classical statistical mechanics: in 1D systems no finite temperature phase transition exists, and the order parameter becomes nonzero only at zero temperature~\cite{nishimori2011elements}.
As we will show, for SWSSB crossover the Krylov complexity remains non-singular.

To illustrate the connection between decoherence-induced phenomena and Krylov space/complexity, we begin with the decohered Ising model with nearest-neighbor dephasing, which exhibits a crossover, and then proceed to the decohered Ising model with infinite-range dephasing, which undergoes a genuine SWSSB phase transition. The conceptual connection between Krylov complexity and conventional SSB is clarified in End Matter.

\textit{Ising Model with Nearest-Neighbor Dephasing Channel---}We consider a 1D chain of $L$ qubits, initialized in $\rho_{\text{init}}=\vert\psi_{\text{init}}\rangle\langle\psi_{\text{init}}\vert$ with all spins aligned along the $X$-direction, i.e. $\vert\psi_{\text{init}}\rangle=\prod_{i=1}^{L}\vert X_i=+1\rangle$. The system then evolves under a nearest-neighbor dephasing channel, $\mathcal{E}[\rho]=\prod_{i=1}^{L-1}\mathcal{E}_{i}[\rho]$, with
\begin{equation}
\mathcal{E}_{i}[\rho]=(1-p)\rho+pZ_iZ_{i+1}\rho Z_iZ_{i+1},
\label{eq:nn-decohere}
\end{equation}
where $Z_i$ is the Pauli-$Z$ operator at site $i$. The decoherence strength $p$ lies in the range $0\leq p \leq 1/2$, with $p=1/2$ corresponding to maximal decoherence. Importantly, this nearest-neighbor dephasing channel preserves both strong and weak $\mathbb{Z}_2$ symmetry generated by $U_{\mathbb{Z}_2}=\prod_{i=1}^{L} X_i$~\cite{see-supplemental}.

Using the double Hilbert space formalism~\cite{see-supplemental}, this channel can be represented as an operator acting on a purified state: the density matrix $\rho=\sum_{\alpha\beta}\rho_{\alpha\beta}\vert \alpha\rangle\langle \beta\vert$ is mapped to a vector $\vert\rho\rangle = \sum_{ij}\rho_{ij}  \vert i\rangle_{u}  \vert j^{*}\rangle_{\ell}$,
where subscriptions $u$ and $\ell$ denote the upper and lower layers of the double Hilbert space. In this representation, the dephasing channel acquires a matrix form in the basis $\vert i\rangle_{u}\vert j^{*}\rangle_{\ell}$, which can be recast as follows~\cite{see-supplemental},
\begin{align}
\mathcal{E}\vert \rho\rangle&=
\prod_{i=1}^{L-1}\left[(1-p) I_{i}^{u}I_{i+1}^{u}I_{i}^{\ell}I_{i+1}^{\ell}  +  pZ_{i}^{u}Z_{i+1}^{u}Z_{i}^{\ell}Z_{i+1}^\ell\right]
\vert\rho\rangle  \nonumber \\
&=e^{-(L-1)\tau}e^{-\tau H^{\text{NN}}}\vert\rho\rangle,
\label{eq:nn-decohere-double}
\end{align}
where $H^{\text{NN}}=-\sum_{i=1}^{L-1}Z_{i}^{u}Z_{i+1}^{u} Z_{i}^{\ell}Z_{i+1}^{\ell}$ and $\tau = -\left[\ln{(1-2p)}\right]/2$. This result suggests that the nearest-neighbor dephasing channel can be regarded as an imaginary-time evolution, with $H^{\text{NN}}$ as an effective Hamiltonian and $\tau$ as the imaginary time, i.e. $\vert\rho(\tau)\rangle\sim e^{-\tau H^{\text{NN}}}\vert\rho_{\text{init}}\rangle$.
Under the time evolution, the decohered system approaches the ground state of $H$, which may exhibit an SWSSB transition at a finite critical time $\tau_c$ or merely a crossover ($\tau_c\to\infty$), as in this model. It is crucial to distinguish these two scenarios. As we will show below, the Krylov complexity serves as a sharp and reliable diagnostic.

\textit{Lanczos Coefficients and Krylov Complexity---}The standard Krylov formalism~\cite{Lanzcos-Book} starts from an initial state $\vert\Psi_{\text{init}}\rangle$ and a time evolution with $\vert\Psi(t)\rangle = e^{-iHt}\vert\Psi_{\text{init}}\rangle$. The corresponding Krylov space is spanned by $\{H^{n}\vert\Psi\rangle\}$. Applying the Gram--Schmidt procedure recursively to $\{H^{n}\vert\Psi\rangle\}$ generates the orthonormal Krylov basis $\{\vert K_{n}\rangle\}$, which satisfies the standard three-term recurrence relation~\cite{Lanzcos-Book},
\begin{align}
    \vert V_{n+1}\rangle = (H-a_{n})\vert K_{n}\rangle - b_{n}\vert K_{n-1}\rangle,\,
    \vert V_{n}\rangle = b_{n}\vert K_{n}\rangle\label{eq:recurrence_eq}
\end{align}
where the Lanczos coefficients $a_{n}$ and $b_{n}$ are defined as
\begin{align}
    a_{n} = \langle K_{n}\vert H\vert K_{n}\rangle,\quad
    b_{n} = \langle V_{n}\vert V_{n}\rangle^{1/2},
\end{align}
with $b_{0} = 0$ and $\vert K_{0}\rangle = \vert\Psi_{\text{init}}\rangle$.
If $b_{n}$ vanishes at some $n\neq0$, the recursive procedure terminates.
Crucially, when expressed in the Krylov basis, the Hamiltonian is always tridiagonal according to Eq.~(\ref{eq:recurrence_eq}), forming an effective 1D tight-binding model, irrespective to the system dimensionality. Furthermore, the time-dependent state in the Krylov basis becomes $\vert\Psi(t)\rangle = e^{-iHt}\vert K_{0}\rangle = \sum_{n}\psi_{n}(t)\vert K_{n}\rangle$, where the expansion coefficient is given by $\psi_{n}(t)=\langle K_{n}\vert e^{-iHt}\vert K_{0}\rangle$ and can be directly evaluated.

In this work, however, we focus on the Krylov complexity of the imaginary-time evolution of the decohered state in Eq.~(\ref{eq:nn-decohere-double})---here $it$ and $\vert\Psi_{\text{init}}\rangle$ are replaced by $\tau$ and $\vert\rho_{\text{init}}\rangle$, respectively, giving $\vert\rho(\tau)\rangle=\sum_{n}\psi_{n}(\tau)\vert K_{n}\rangle$.
To substantially simplify the notation in the following, although this is not required for the evolution itself, we impose the normalization condition $\langle\rho(\tau)\vert\rho(\tau)\rangle=1$ at each $\tau$, ensuring $\sum_{n}\abs{\psi_{n}(\tau)}^{2}=1$.
The Krylov complexity for decohered systems is then defined as
\begin{equation}
\mathcal{K}(\tau)=\sum_n n\abs{\psi_n(\tau)}^2.
\label{eq:K-complexity}
\end{equation}

\begin{figure}[t]
\begin{center} 
\includegraphics[width=6.5cm]{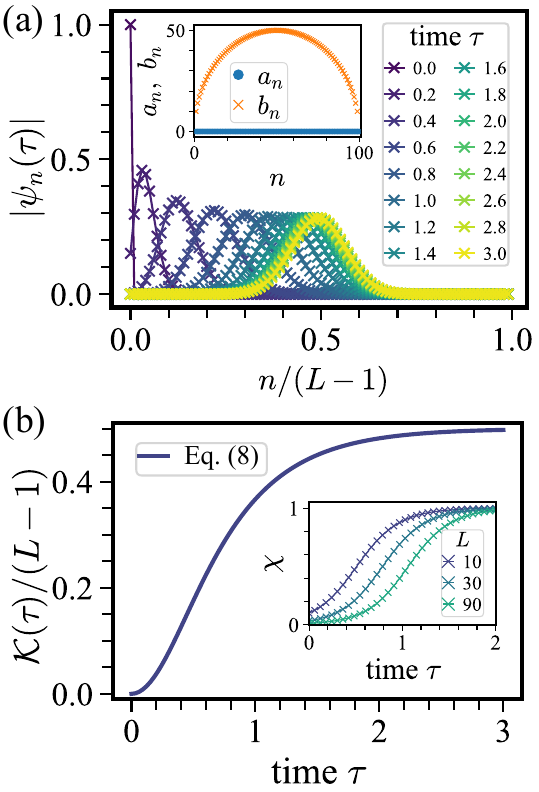}  
\end{center} 
\vspace{-.3cm}
\caption{(a) Time evolution of the wave-packets in the Krylov basis for $H^{\text{NN}}$ which exhibits an SWSSB crossover. The wave packet spreads smoothly and develops a Gaussian form as $\tau$ increases. The inset shows the corresponding Lanczos coefficients; $L=100$. (b) Normalized Krylov complexity $\mathcal{K}(\tau)/(L-1)$ for $H^{\text{NN}}$, displaying a nonsingular profile that indicates the absence of a phase transition. The inset shows the time evolution of the R{\'e}nyi-2 correlator $\chi$ for different system sizes, further confirming the crossover behavior.}
\label{Fig2}
\end{figure}

While $\mathcal{K}(t)$ characterizes the spread of $\vert\Psi(t)\rangle$ within Hilbert space, $\mathcal{K}(\tau)$ quantifies the ``information loss’’ of $\vert\rho_{\text{init}}\rangle$ under decoherence. This perspective follows from the structure of the Krylov basis ${\vert K_{n}\rangle}$, where each state represents precisely $n$ applications of the error operator. Specifically, in the Gram--Schmidt procedure,  $\vert K_{0}\rangle=\vert\rho_{\text{init}}\rangle$ is the no-error state, $b_{1}\vert K_{1}\rangle = H\vert K_{0}\rangle - \vert K_{0}\rangle\langle K_{0}\vert H\vert K_{0}\rangle$ is the one-error state with the no-error contribution projected out, and more generally $b_{n+1}\vert K_{n+1}\rangle = H\vert K_{n}\rangle - \sum_{m=0}^{n}\vert K_{m}\rangle\langle K_{m}\vert H\vert K_{n}\rangle$ corresponds to the $(n+1)$-error state with all lower-error contributions removed, as illustrated schematically in Fig.~\ref{Fig1} (b). We remark that $\mathcal{K}(\tau)$ gives the average number of noise events applied to $\vert\rho_{\text{init}}\rangle$, measuring the degree of decoherence in $\vert\rho(\tau)\rangle$.

The Lanczos coefficients and Krylov complexity of this model can be obtained analytically. The effective Hamiltonian $H^{\text{NN}}$ commutes with a local parity symmetry $g_{i}\equiv X_{i}^{u}X_{i}^{\ell}$, and the initial state $\vert\rho_{\text{init}}\rangle=\prod_{i=1}^{L}\vert X_{i}=+1\rangle_{i}^{u}\vert X_{i}=+1\rangle_{i}^{\ell}$ is invariant under all $g_{i}$. This allows us to restrict the analysis to the positive-parity sector, reducing the Hilbert-space dimension from $4^{L}$ to $2^{L}$. A convenient basis for this sector is given by $\vert\Uparrow_{i}\rangle=\left( \vert\uparrow^{u}_{i}\uparrow^{l}_{i}\rangle + \vert\downarrow^{u}_{i}\downarrow^{l}_{i}\rangle\right)/\sqrt{2}$ and $\vert\Downarrow_{i}\rangle=\left(\vert\uparrow^{u}_{i}\downarrow^{l}_{i}\rangle+\vert\downarrow^{u}_{i}\uparrow^{l}_{i}\rangle\right)/\sqrt{2}$, in which the Hamiltonian takes the classical Ising form $H^{\text{NN}}=-\sum_{i=1}^{L-1}\tau^{z}_{i}\tau^{z}_{i+1}$, where $\tau^{z}_{i}$ represents the Pauli-$Z$ operator at site $i$, and $\vert\rho_{\text{init}}\rangle = \prod_{i=1}^{L} \left(\vert\Uparrow_{i}\rangle+\vert\Downarrow_{i}\rangle\right)/\sqrt{2} = \prod_{i=1}^{L}\vert \tau_{i}^{x}=+1\rangle$.

We then apply the Kramers--Wannier (KW) transformation to this open-chain Hamiltonian~\cite{PhysRevB.108.214429}, obtaining $H^{\text{NN}} = -\sum_{i=1}^{L-1}\tau^{x}_{i+1/2} \equiv -2S^{x}=-(S^{+}+S^{-})$ where $\tau^{x}_{i+1/2}$ denotes the Pauli-$X$ operator on the link $i+1/2$ and $S^{x}$ represents the collective spin operator satisfying the $\mathfrak{su}(2)$ algebra. The Lanczos coefficients can be calculated analytically by the algebraic method~\cite{see-supplemental}: we first express the Hamiltonian in the spin basis $\{\vert s,m\rangle\}$, in which it becomes tridiagonal. Using the mapping $\vert s=(L-1)/2,m=s-n\rangle\rightarrow (-1)^n \vert K_n\rangle$, with $\vert\rho_{\text{init}}\rangle=\prod_{i=1}^{L-1}\vert \tau_{i+1/2}^{z}=+1\rangle=\vert K_{0}\rangle$ corresponding to the highest-weight state, the Hamiltonian takes the form of the three-term recurrence relation, Eq.~(\ref{eq:recurrence_eq}). From this comparison, the Lanczos coefficients are read off as
\begin{align}
    &a_{n}=0;\quad
    b_{n}=\sqrt{n(L-n)}.
\end{align}
While the amplitude $\psi_{n}(\tau)$, and $\mathcal{K}(\tau)$ can, in general, be formulated using the coherent-state approach~\cite{Coherent_state1,Coherent_state2}, here we directly evaluate the time evolution by applying the Baker--Campbell--Hausdorff formula, resulting in
\begin{align}
    &\psi_{n}(\tau) =(-1)^n\sqrt{{L-1\choose n}\lambda^n(1-\lambda)^{L-1-n}},\label{eq:target1-psi}\\
    &\mathcal{K} = (L-1)\lambda,\label{eq:K_NN}
\end{align}
where $\lambda=\frac{\sinh^2(\tau)}{1+2\sinh^2(\tau)}$. Details are provided in the Supplemental Material~\cite{see-supplemental}.

Figure~\ref{Fig2} (a) shows the evolution of ${\psi}_n(\tau)$, with the inset illustrating the Lanczos coefficients.
As time progresses, the wave packet spreads, indicating increasing decoherence of the quantum state. The emergence of a Gaussian wave packet can be attributed to the binomial distribution in Eq.~(\ref{eq:target1-psi}): in the large $L$ limit, $\abs{\psi_{n}}^{2}$ approaches a normal distribution with mean $L\lambda$ and variance $L\lambda(1-\lambda)$. Physically, the evolution operator associated with the KW transformed Hamiltonian acts as independent spin-flip operations on the $L-1$ link spins. Hence, $\abs{\psi_{n}(\tau)}^2$ which corresponds to the probability of finding $n$ flipped spins (equivalently $n$ errors) with a given flipping probability---controlled by $\lambda$---naturally flows to the binomial form in Eq.~(\ref{eq:target1-psi}).

Figure~\ref{Fig2} (b) displays $\mathcal{K}(\tau)/(L-1)$. As $\tau$ increases, $\mathcal{K}(\tau)/(L-1)$ rises monotonically and approaches $0.5$ without exhibiting any singular behavior, indicating the absence of a phase transition at finite $\tau$. This smooth behavior suggests that $\vert\rho\rangle$ is susceptible to errors and gradually evolves toward the SWSSB state. To further confirm the absence of an SWSSB phase transition, we employ the tensor-network technique~\cite{Tensor-network1,Tensor-network2,see-supplemental} to compute the R\'{e}nyi-2 correlator, $\chi=\frac{1}{L^2}\sum_{ij}\langle \rho\vert Z_{i}^{u}Z_{i}^{\ell}Z_{j}^{u}Z_{j}^{\ell}\vert\rho\rangle/\langle \rho\vert\rho\rangle$.
The inset shows the evolution of $\chi$, which shows a non-crossing behavior as a function of $L$, confirming that the system undergoes a crossover rather than a sharp SWSSB phase transition.

\textit{Ising Model with Infinite-Range Dephasing Channel---}We now turn to an example where SWSSB genuinely occurs. Starting from the same initial state $\vert\psi_{\text{init}}\rangle=\prod_{i=1}^{L}\vert X_i=+1\rangle$, we apply an infinite-range dephasing channel studied in Ref.~\cite{IR-decohere}---namely, instead of restricting the $ZZ$ dephasing to nearest-neighbor sites, we apply it between all pairs of sites.
For simplicity, we present $\mathcal{E}[\rho]$ only as an imaginary-time evolution in the double Hilbert space formalism; details are provided in \cite{see-supplemental}:
\begin{align}
\vert\mathcal{E}[\rho]\rangle=e^{-\tau H^{\text{IR}}}|\rho\rangle,
\label{eq:IR-decohere-double}
\end{align}
where $H^{\text{IR}}=-\sum_{i<j} \left(Z_i^u Z_i^l Z_j^u Z_j^l -1 \right)/L$, and $0\le\tau<\infty$.
In this case, an SWSSB phase transition is expected to occur at $\tau_{\text{c}}=0.5$~\cite{IR-decohere}.

The effective Hamiltonian $H^{\text{IR}}$ also commutes with the local parity symmetry $g_{i}$. Consequently, we again reduce the Hilbert space by restricting to the positive-parity sector. In this reduced space, the Hamiltonian takes the the infinite-range Ising form $H^{\text{IR}} = -\sum_{i<j}\left(\tau^{z}_{i}\tau^{z}_{j}-1\right)/L$.

While up to this point all Krylov quantities can be obtained numerically for small systems ($L< 16$)~\cite{systemsize,Alishahiha2025}, we now provide analytical solutions for the Lanczos coefficients (see \cite{see-supplemental} for details), which enable access to larger sizes ($L\sim 500$). We introduce the collective spin operator $S_{z}\equiv\sum_{i}\tau^{z}_{i}/2$, allowing the Hamiltonian to be written as $H^{\text{IR}}=-\frac{2}{L}S_{z}^{2}+\frac{L}{2}$. Applying a spin rotation $\mathcal{R}_{i}=e^{-i\pi\tau^{y}_{i}/4}$ exchanges $x\rightarrow z\rightarrow -x$, transforming the Hamiltonian into the Lipkin--Meshkov--Glick form $H^{\text{IR}}=-\frac{2}{L}S_{x}^{2}+\frac{L}{2}$~\cite{LMG-1,LMG-2,LMG-3,LMG-4}.
The $\mathfrak{su}(2)$ algebra again provides a natural framework for deriving analytic expressions for the Lanczos coefficients. Proceeding as before, the Hamiltonian is tridiagonal in the spin basis, and under the mapping $\vert s=L/2,m=L/2-2n\rangle\mapsto(-1)^n\vert K_{n}\rangle$ one reads off,
\begin{align}
    &a_{n} = -2n + \frac{4}{L}n^{2} - \frac{1}{2} + \frac{L}{2},\label{eq:H_IR_an}\\
    &b_{n} = \frac{1}{2L} \sqrt{2n (L-2n+1) (2n-1) (L-2n+2)},\label{eq:H_IR_bn}
\end{align}
which enables us to perform exact diagonalization for large-size Hamiltonians.

\begin{figure}[t]
\begin{center} 
\includegraphics[width=6.5cm]{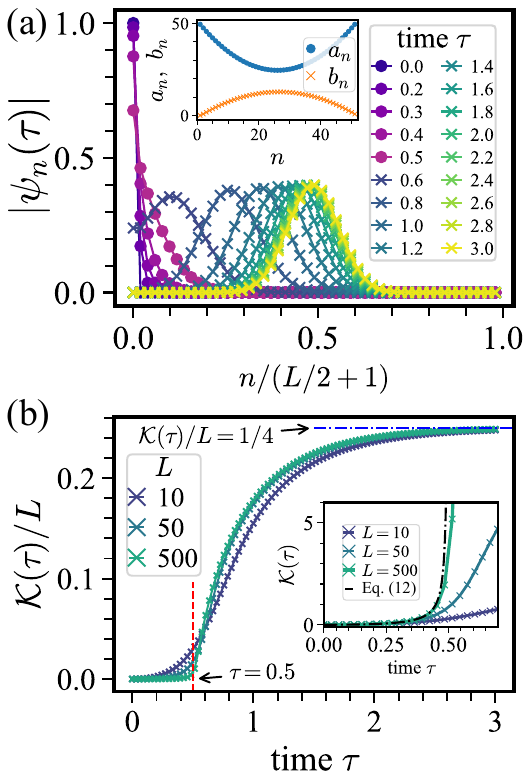}  
\end{center} 
\vspace{-.3cm}
\caption{(a) Time evolution of the wave packets in the Krylov basis for $H^{\text{IR}}$, which exhibits a genuine SWSSB phase transition. The packet remains localized within the low-error subspaces for $\tau\le0.5$ (dotted lines), and begins to spread when $\tau$ exceeds the critical point $\tau=0.5$. The inset shows the corresponding Lanczos coefficients; $L=100$. (b) Normalized Krylov complexity $\mathcal{K}(\tau)/L$ for $H^{\text{IR}}$. In the thermodynamic limit ($L=500$), the Krylov complexity shows a sharp transition at $\tau=0.5$. $\mathcal{K}$ follows an area-low scaling ($\mathcal{K}\sim\mathrm{constant}$) for $\tau<0.5$ and a volume-law scaling ($\mathcal{K}\sim\mathcal{O}(L^{1})$) in the long time limit.
The inset compares numerical (colored) and analytical result in the thermodynamic limit (black) in the area-law regime, confirming excellent agreement.}
\label{Fig3}
\end{figure}

Figure~\ref{Fig3} (a) shows the evolution of $\psi_n(\tau)$. Unlike the nearest-neighbor case, where $\psi_{n}(\tau)$ simply broadens continuously, $\psi_{n}(\tau)$ of the Ising model with the infinite-range dephasing channel remains strongly localized around the small-error states up to $\tau=0.5$, and begins to spread only for $\tau>0.5$, forming a Gaussian profile. The early-time localization ($\tau<0.5$) implies that $\psi_n(\tau)$ is robust against errors, whereas the subsequent localization-delocalization transition marks the onset of the SWSSB phase transition.

To further clarify this correspondence, we numerically compute $\mathcal{K}(\tau)$.
Figure~\ref{Fig3} (b) displays the evolution of $\mathcal{K}(\tau)/L$. In the thermodynamic limit, $\mathcal{K}(\tau)/L$ remains zero for $\tau<0.5$, indicating an area-law scaling of the Krylov complexity. In contrast, once $\tau$ exceeds $0.5$, $\mathcal{K}(\tau)/L$ undergoes a sharp transition and saturates to a finite constant at long times, signifying that the Krylov complexity exhibits a volume-law scaling $\mathcal{K}(\tau)\propto L$. This abrupt change at $\tau=0.5$ unambiguously identifies the SWSSB phase transition.

Both the area-law scaling at small $\tau$ and the volume-law scaling at large $\tau$ can be analyzed analytically in the thermodynamic limit ($L\rightarrow\infty$). Here we outline the derivations, with full details provided in \cite{see-supplemental}. Our goal is to compute $\psi_n(\tau) = \langle K_{n}\vert \rho(\tau)\rangle/\langle\rho(\tau)\vert\rho(\tau)\rangle$ where $\vert\rho(\tau)\rangle = e^{-H^{\text{IR}}\tau}\vert K_{0}\rangle$. For the area-law regime, we apply the Holstein--Primakoff transformation~\cite{PhysRev.58.1098} to the spin operators.
Expanding $H^{\text{IR}}$ to leading order in spin yields an accurate description in the thermodynamic limit. In this limit, $\psi_{n}(\tau)$ reduces to integrals over Hermite polynomials, which can be solved analytically for $0\le \tau<1/2$, giving
$\psi_n(\tau) = \frac{\sqrt{(2n)!}}{2^{n}n!}\sqrt{\frac{1}{1-\tau}}\left(\frac{-\tau}{1-\tau}\right)^{n}/\left(\frac{1}{1-2\tau}\right)^{1/4}$. For large $n$, this becomes an exponentially localized state $\psi_n\propto\frac{(-1)^n}{(\pi n)^{1/4}}e^{-n\ln{(\frac{1-\tau}{\tau})}}$. Importantly, $\psi_n(\tau)$ exhibits no $L$ dependence within the domain $0\le\tau<1/2$, ensuring an area-law profile for the Krylov complexity,
\begin{align}
    \mathcal{K}(\tau) = \frac{\tau^{2}}{2(1-2\tau)}.\label{eq:area_law_K}
\end{align}
In the inset of Fig.~\ref{Fig3} (b), we compare this analytic result with numerical evaluations of $\mathcal{K}$ for $0<\tau<1/2$, finding agreement for large $L$.

For the volume-law regime, we adopt an alternative approach to analyze the asymptotic behavior in the limit of large $\tau$ and $L$ with fixed $\tau/L$. We express $\psi_n(\tau)$ in terms of Wigner $d$-matrices~\cite{sakurai2020modern}. In the large-$\tau$, large-$L$ limit, the dominant contributions arise from the high-spin components, yielding $\psi_n(\tau)  = (-1)^n\sqrt{{L \choose 2n}2^{1-L}}$, which is the square root of binomial distribution and again approaches a Gaussian distribution for large $L$.
The corresponding Krylov complexity is then
\begin{align}
    \mathcal{K} = \frac{L}{4},\label{eq:K_complexity_volume_law}
\end{align}
which follows the volume law and agrees with our numerical results, as illustrated in Fig.~\ref{Fig3} (b).

\textit{Summary and Outlook---}In this Letter, we generalize the concept of Krylov complexity to decohered systems.
Through both numerical and analytical analyses, we demonstrate that Krylov complexity provides an effective probe for detecting SWSSB phase transitions intrinsic to mixed states.
Furthermore, we point out that the spreading of the wave packet in the Krylov subspace reflects how intricately errors proliferate in the quantum state---an effect naturally quantified by Krylov complexity.
As clarified in the End Matter, the connection between Krylov complexity and SWSSB can also be interpreted in terms of an effective free-energy description. Our results thus offer a new complexity-based perspective on mixed state phase transitions.

A natural next step is to apply this framework to phase transitions of intrinsic mixed topological orders~\cite{MixedTO1,MixedTO2,MixedTO3,MixedTO4} and average symmetry-protected topological phases~\cite{ASPT1,ASPT2,ASPT3,ASPT4}.
In addition, while we confirm that Krylov complexity quantifies information loss of initial states, its relationship to the recoverability of the states~\cite{recoverbility-1,recoverability-2,recoverability-3,recoverability-4,QEC-futu,Information,Information2} remains unexplored.
Clarifying this connection with information-theoretic quantities that characterize the recoverability of the initial states would be an important future direction.
Another promising avenue is to investigate
whether decohered systems exhibit an analogue of operator spreading~\cite{Operator-spreading,OTOC,Strong_zero_mode1,Strong_zero_mode2,PhysRevX.8.021014}, a central theme in Krylov complexity studies, with diagnostics provided by R{\'e}nyi-1, R{\'e}nyi-2, fidelity, and Wightman correlators.

\textit{Acknowledgments---}
This work was supported by JST PRESTO (Grant No. JPMJPR2359) and JSPS KAKENHI (Grant No. 24H00829) (H.-H.T.), and by JSPS KAKENHI (Grant No. JP23KJ0360 and JP26K17056) (T.O.).\\

\bibliography{apssamp}

\newpage

\begin{center}
\bf\large End Matter
\end{center}

In the main text, we unify the framework of decoherence and Krylov complexity, and show that the Krylov complexity provides a useful probe of mixed-state phase transitions, including the regime where SWSSB occurs. In this End Matter, we clarify why this is so by relating the Krylov construction to an effective free-energy description familiar from the Landau paradigm. This connection does not replace the Krylov picture, but explains why Krylov observables are naturally sensitive to phase transitions and SSB.

\textit{General Framework---}A useful starting point is the return amplitude (or survival amplitude),
\begin{align}
r(\tau)=\langle\rho(0)\vert\rho(\tau)\rangle =
\langle\rho(0)\vert e^{-H\tau} \vert\rho(0)\rangle,
\end{align}
where $\vert\rho(\tau)\rangle$ is an unnormalized imaginary-time-evolved state.
On the one hand, $r(\tau)$ serves as a generating function for the moments $\mu_{n} \equiv d^{n}r / d\tau^{n} \vert_{\tau=0}$, from which the Lanczos coefficients $a_{n}$ and $b_{n}$ and hence the Krylov complexity are determined\cite{Lanzcos-Book,Operator-spreading,BalasubramanianPRD2022}. On the other hand, the same quantity defines an effective free energy $F$\cite{SWSSB1},
\begin{align}
F(\tau)=-\log[\langle\psi(\tau)|\psi(\tau)\rangle]= -\log r(2\tau).
\end{align}
Here $r(2\tau)$ plays the role of an effective partition function, and so in the following we denote it as $Z(\tau) = r(2\tau)$. We remark that in a suitable thermodynamic or large-deviation limit, $F(\tau)$ encodes the singularity structure of phase transitions and supports a Landau-type description. Therefore, the return amplitude provides a useful bridge:
\begin{align}
    \mathcal{K} \leftrightarrow
    a_{n},b_{n} \leftrightarrow
    \mu_{n} \leftrightarrow
    r \leftrightarrow
    F.\label{eq:bridge_complexity_and_free_energy}
\end{align}
This correspondence suggests that the nonanalyticities in $F(\tau)$ must be encoded in the Krylov hierarchy. Likewise, SSB structure in the effective free-energy description should leave observable imprints in the Krylov quantities as well. In fact, part of the connection in Eq. (\ref{eq:bridge_complexity_and_free_energy}) (between $\mathcal{K}$ and $r$) can already be seen intuitively from the localization-to-delocalization structure discussed in the main text. In the area law regime, the state $\vert\rho(\tau)\rangle$ remains concentrated in low-error sectors, so $\mathcal{K}$ stays small and the state retains a large overlap with $\vert\rho(0)\rangle$, leading to large $r$; in the volume law regime, the state spreads into higher-error sectors, causing $\mathcal{K}$ to grow while the return amplitude is strongly suppressed.

In the following, we show how this correspondence is realized in the long-range dephasing $H^{\text{IR}}$ studied in the main text.

\textit{Long-Range Dephasing Channel} $H^{\text{IR}}$---
We begin with the effective partition function associated with $H^{\text{IR}}$,
\begin{align}
Z(\tau)=&\langle\rho(0)|e^{-2\tau H^{\text{IR}}}|\rho(0)\rangle\nonumber\\
=&e^{-L\tau/2} \sum_{m'=-\frac{L}{2}}^{\frac{L}{2}} {L \choose \frac{L}{2}+m'} 2^{-L} e^{4m'^{2}\tau/L}\notag\\
=&\sum_{q=-1}^{1} {L \choose \frac{L}{2}(1+q)} 2^{-L} e^{L\tau q^{2}} e^{-L\tau/2},
\end{align}
where $m'$ labels the eigenvalue of $S_x$, and we introduce a magnetization $q\equiv\frac{2m'}{L}\in[-1,1]$, which will serve as an order parameter in the following. The last line naturally decomposes $Z(\tau)$ into contributions from magnetization sectors labeled by $q$, and we denote the summand by $Z(\tau,q)$.
In the large-$L$ limit, each sector contribution $Z(\tau,q)$ takes the large-deviation form,
\begin{align}
Z(\tau,q)\asymp e^{-L\phi_\tau(q)},
\end{align}
where $\phi_\tau(q)$ plays the role of a constrained free-energy density~\cite{LDA1,LDA2}. Using the Stirling's formula for the binomial coefficient in $Z(\tau,q)$, we obtain
\begin{align}
L\phi_\tau(q)=&-\ln[Z(\tau,q)]\nonumber\\
\simeq& L\biggl[\frac{1}{2}[(1+q)\ln(1+q)\notag\\
&+(1-q)\ln(1-q)]- \tau q^2 -\frac{\tau}{2}\bigg]
\end{align}

We first focus on the small $\tau$ case, where the dominant saddle lies near $q=0$:
\begin{align}
\phi_\tau(q)\simeq -\frac{\tau}{2} + \left( \frac{1}{2}-\tau \right) q^2+\frac{1}{12}q^4+\mathcal{O}(q^6)\label{eq:Landauene-IR},
\end{align}
This has the standard Landau form---the quadratic coefficient changes sign at $\tau_{\text{c}}=1/2$ and $\phi_{\tau}(q)$ becomes a Mexican-hat shape, indicating an SSB when $\tau>\tau_{\text{c}}$.

The order parameter $q$ at arbitrary $\tau$ is determined by the stationary condition $\frac{\partial \phi_\tau(q)}{\partial q}=0$, leading to
\begin{align}
\frac{1}{2}\ln\left[\frac{1+q}{1-q}\right]=2\tau q \Rightarrow q=\tanh(2q\tau).
\end{align}
The stability of a given saddle is determined by
\begin{align}
\frac{\partial^2 \phi_\tau(q)}{\partial q^2}=\frac{1}{1-q^2}-2\tau,\label{eq:stability-saddle}
\end{align}
which depends on both $\tau$ and $q$.
At the symmetric saddle $q=0$, the curvature changes sign at $\tau_c=1/2$, indicating loss of stability.
As a result, for $\tau>1/2$, the minima move to nonzero $q$, and the system enters the (SW)SSB phase.

We then make the connection between the Krylov complexity $\mathcal{K}$ and the (SW)SSB explicit. In conventional SSB, such as quantum Ising model, the ordered phase is characterized by the long range order. A useful diagnostic is the fluctuation of the order operator ($S_x=\sum_i \tau^x_i$ in this case), which vanishes in the disordered phase and remains finite in the ordered phase~\cite{Tasaki_text}. In particular, in 1D systems with a finite length $L$, the diagnostic has the following form:
\begin{align}
    \sqrt{
    \left\langle \left(\frac{S_x}{L}\right)^2 \right\rangle
    - \left\langle \frac{S_x}{L} \right\rangle^2
    }.
\end{align}
For $H^{\text{IR}}$, the second term vanishes by the selection rule: $S_x$ changes the relevant spin quantum number by one unit, whereas the Krylov basis has supported only on basis states whose quantum numbers differ by two units. The remaining first term can be evaluated straightforwardly and is directly related to the Krylov complexity~\cite{see-supplemental}. In the area law regime,
\begin{align}
    \frac{\mathcal{K}}{L} = 2\tau^2 \left\langle \left(\frac{S_x}{L}\right)^2 \right\rangle = \frac{\tau^2}{2(1-2\tau)L},\label{eq:connection_K_Sx2_area}
\end{align}
which approaches zero in the thermodynamic limit, while in the volume-law regime with large $\tau$ and large $L$ limits,
\begin{align}
    \frac{\mathcal{K}}{L} = \left\langle \left(\frac{S_x}{L}\right)^2 \right\rangle = \frac{1}{4}\label{eq:connection_K_Sx2_volumn}
\end{align}
Thus, the Krylov complexity directly captures the fluctuation of $S_x$, and hence probes the onset of (SW)SSB. 

Lastly, we remark that, unlike the complexity-based diagnosis, this Landau-potential analysis relies on identifying a suitable constrained free-energy description or an order parameter in advance. To illustrate this point, consider instead the nearest-neighbor dephasing channel $H^{\text{NN}} = -\sum_{i=1}^{L-1} \tau_{i+1/2}^{x}$ in its Kramers--Wannier dual form. If one naively repeats the above construction using the magnetization in the $x$ direction as the order parameter $q$, one obtains:
\begin{align}
Z(\tau)=&\sum_{q=-1}^{1} {L-1 \choose \frac{L-1}{2}(1+q)} 2^{-(L-1)} e^{2\tau (L-1)q},\\
L\phi_\tau(q) \simeq& L\biggl[\frac{1}{2}[(1+q)\ln(1+q)\notag\\
&+(1-q)\ln(1-q)]- 2\tau q\biggr].
\end{align}
Let us focus on the small $\tau$ limit, where the dominant saddle lies near $q=0$, so $\phi_{\tau}(q)$ becomes
\begin{align}
\phi_\tau(q)\simeq -2\tau q+\frac{1}{2}q^2+\frac{1}{12}q^4+{\mathcal{O}(q^6)}\label{eq:Landauene-NN}.
\end{align}
The resulting constrained-free-energy density is not symmetric under $q\rightarrow-q$, because of the linear term $-2\tau q$. This shows that, for $H^{\text{NN}}$, the magnetization in the $x$ direction is not the appropriate Landau variable: under the Kramers--Wannier dual, it does not play the role of the conventional symmetry-breaking order parameter associated with the $z$-magnetization, i.e. $\sum_{i=1}^{L-1}\tau_{i+1/2}^{x}\xrightarrow{\text{KW}^{-1}}\sum_{i=1}^{L-1}\tau_{i}^{z}\tau_{i+1}^{z}$.

Therefore, while the effective free-energy construction is useful for making contact with the conventional Landau language, the Krylov framework is more microscopic and exact at the lattice level: the complexity directly measures the spreading of the evolving state in Krylov space, without any coarse graining, and thus remains applicable even when an effective Landau potential or an appropriate order parameter is not evident.

\onecolumngrid
\newpage

\renewcommand{\thesection}{\Alph{section}}
\renewcommand{\theequation}{\thesection\arabic{equation}}
\renewcommand{\thefigure}{\thesection\arabic{figure}}

\makeatletter
\@addtoreset{equation}{section}  

\setcounter{section}{0}
\setcounter{equation}{0}
\setcounter{figure}{0}
\setcounter{page}{1}

\makeatletter
\let\SM@sec@upcase\sec@upcase
\let\SM@MakeTextUppercase\MakeTextUppercase
\def\sec@upcase#1{#1}
\renewcommand{\MakeTextUppercase}[1]{#1}
\makeatother

\begin{center}
\bf\large Supplemental Material --- Krylov Complexity and Mixed-State Phase Transition
\end{center}

\begin{center}
Hung-Hsuan Teh\\
\textit{The Institute for Solid State Physics, The University of Tokyo, Kashiwa,
Chiba, 277-8581, Japan}\\
\textit{Graduate School of Informatics, Nagoya University, Nagoya, 464-0814, Japan}
\end{center}

\begin{center}
Takahiro Orito\\
\textit{Department of Physics, College of Humanities and Sciences, Nihon University, Sakurajosui, Setagaya, Tokyo 156-8550, Japan}
\end{center}

\section{Strong and Weak Symmetries}
In this section, we briefly explain two types of symmetries in the density matrix: strong and weak symmetries. We also discuss the symmetry classification of the quantum channels. For more details, see \cite{Buca_2012,Albert_2014,groot2022,Ex1,symmetry-in-OQS}.

The density matrix $\rho$ possesses two distinct symmetries. $\rho$ exhibits strong symmetry if $U_{g}\rho=e^{i\theta}\rho$, where $U_{g}$ is a representation of an element $g$ of a symmetry group $G$.
The condition of strong symmetry requires that all eigen vectors of the density matrix, denoted as $|\lambda_i\rangle$, remain invariant under the action of $U_{g}$, i.e., $U_{g}|\lambda_i\rangle=e^{i\theta}|\lambda_i\rangle$,
with $\theta$ being a single phase.
In other words, all eigenvectors carry the same conserved charge associated with symmetry $G$, similar to a symmetric quantum state for a pure state.
In contrast, $\rho$ exhibits weak symmetry if $U_{g}\rho U_{g}^{\dagger}=\rho$. In this case, all eigen vectors of the density matrix do not have to carry the same conserved charge. Specifically, the density matrix takes a block-diagonal form, with each block corresponding to a different charge sector.

We discuss the conditions under which the quantum channel preserves strong/weak symmetry.
Here, we utilize the operator-sum representation of the channel, as described in
\begin{align}
\mathcal{E}[\rho] = \sum_{m}K_{m}\rho K_{m}^{\dagger},
\end{align}
where $K_m$ represents a set of Kraus operators that satisfy $\sum_m K_m^\dagger K_m=I$, with $I$ denoting the identity matrix.

The condition for a channel to preserve strong symmetry is given by
\begin{align}
K_{m}U_g= U_gK_{m}\ \ \forall m,\ \forall g\in G.
\end{align}
That is, $K_m$ commutes with $U_g$.
On the other hand,
the condition for a channel to preserve weak symmetry is given by
\begin{align}
U_g\left[\sum_m K_{m}\rho K_m^\dagger\right]U_g= \sum_m K_{m}\rho K_m^\dagger\quad\forall g\in G,
\end{align}
or alternatively,
\begin{align}
K_{m}U_g= e^{i\phi_m(g)}U_gK_{m},
\end{align}
where $e^{i\phi_m(g)}$ cannot be eliminated by the gauge transformation~\cite{Ex1,symmetry-in-OQS}.

\section{Mapping from Decoherence Channel to Imaginary Time Evolution}\label{sm:decoherence_channel_to_imaginary_t_evolution}
We here construct the mapping from a decoherence quantum channel to an imaginary time evolution, i.e., from Eq.~(\ref{eq:nn-decohere}) to Eq.~(\ref{eq:nn-decohere-double}) in the main text. We begin with the double Hilbert space formalism, which basically reshapes a matrix into a vector irrespective of the specific purification or vectorization method used. In this study, we adopt the mapping $\rho\mapsto\vert\rho\rangle=\sum_{\alpha}\rho \vert\alpha\rangle \otimes \vert\alpha\rangle \equiv \sum_{\alpha}\rho \vert\alpha\rangle_{u} \vert\alpha\rangle_{\ell}$, commonly referred to as the Choi--Jamio\l{}kowski isomorphism. The subscripts $u$ and $\ell$ denote the upper and lower sectors of the doubled Hilbert space.

Given a density matrix in a computational basis $\{\vert i\rangle\}$, which in general may differ from $\{\vert\alpha\rangle\}$, $\rho=\sum_{ij}\rho_{ij} \vert i\rangle \langle j\vert$, and a basis transformation $\vert i\rangle = \sum_{\alpha}c^{\alpha}_{i}\vert \alpha\rangle$, the mapping yields
\begin{align}
    \rho\mapsto
    \vert\rho\rangle &= \sum_{\alpha} \sum_{ij}\rho_{ij}\vert i\rangle\langle j \vert\alpha\rangle\otimes\vert\alpha\rangle\notag\\
    &=\sum_{ij}\rho_{ij}\vert i\rangle \otimes \vert j^{*}\rangle
    =\sum_{ij}\rho_{ij}\vert i\rangle^{u}\vert j^{*}\rangle^{\ell},
\end{align}
where $\vert j^{*}\rangle\equiv\sum_{i}\left(c_{j}^{\alpha}\right)^{*} \vert\alpha\rangle$. As an example, a density matrix $\rho=[\rho_{11}\;\rho_{12};\;\rho_{21}\;\rho_{22}]$ is mapped to the vector $\vert\rho\rangle=[\rho_{11};\;\rho_{21};\;\rho_{12};\;\rho_{22}]$.

We then rewrite a generic quantum channel $\mathcal{E}[\rho] = \sum_{m}B_{m}\rho B_{m}^{\dagger}$, represented by the Kraus operators $\{B_{m}\}$, in the double Hilbert space. Using the identity
\begin{align}
    X\rho Y\mapsto
    \vert X\rho Y\rangle = \left(Y^{\text{T}}\otimes X\right) \vert\rho\rangle,
\end{align}
where $X$ and $Y$ are arbitrary matrices with the same dimension as $\rho$, we obtain
\begin{align}
    \mathcal{E}[\rho]\mapsto
    \vert\mathcal{E}[\rho]\rangle = \sum_{m} B_{m}^{*}\otimes B_{m}\vert\rho\rangle
    =\sum_{m}\left(B_{m}^{u}\right)^{*}B_{m}^{\ell}\vert\rho\rangle.
\end{align}
For the specific quantum channel in Eq.~(\ref{eq:nn-decohere}) of the main text, which involves two Kraus operators $\sqrt{1-p}I_{i}I_{i+1}$ and $\sqrt{p}Z_{i}Z_{i+1}$, this transformation directly yields the first line of Eq.~(\ref{eq:nn-decohere-double}).

To cast the channel in the form of an imaginary time evolution, we employ the following identity,
\begin{align}
    e^{\xi X} = \cosh{\xi} + \left(\sinh{\xi}\right)X,\label{eq:exp_expansion_identity}
\end{align}
where $\xi$ is a scalar and $X$ is an operator satisfying $X^{2}=I$. We also reparameterize $p$ by $\tau=-\left[\ln{(1-2p)}\right]/2$. Accordingly, the quantum channel in the double Hilbert space becomes
\begin{align}
    &(1-p) I_{i}^{u}I_{i+1}^{u}I_{i}^{\ell}I_{i+1}^{\ell}  +  pZ_{i}^{u}Z_{i+1}^{u}Z_{i}^{\ell}Z_{i+1}^\ell\notag\\
    =&e^{-\tau}\left[
    \left(\cosh{\tau}\right)I_{i}I_{i}^{u}I_{i+1}^{u}I_{i}^{\ell}I_{i+1}^{\ell}
    +\left(\sinh{\tau}\right)Z_{i}^{u}Z_{i+1}^{u}Z_{i}^{\ell}Z_{i+1}^\ell
    \right]\notag\\
    =&e^{-\tau}e^{\tau Z_{i}^{u}Z_{i+1}^{u}Z_{i}^{\ell}Z_{i+1}^\ell},
\end{align}
which reproduces the second line of Eq.~(\ref{eq:nn-decohere-double}) in the main body of the text.

\section{Lanczos Coefficients for $H^{\text{NN}}$}
In this section, we compute the Lanczos coefficients for $H^{\text{NN}}$ using its underlying $\mathfrak{su}(2)$ algebraic structure. We also present an alternative derivation based on the relation between Lanczos coefficients and the moment of the correlation function. As a reminder, the effective Hamiltonian derived in the main text reads
\begin{align}
H^{\text{NN}} = -\sum_{i=1}^{L-1}\tau^{x}_{i+1/2} = -(S^{+}+S^{-}),
\label{eq:SU(2) Hamiltonian}
\end{align}
where $\tau^{x}_{i+1/2}$ denotes the Pauli-$X$ operator on the bond $i+1/2$,
and the initial state is the highest-weight state $\vert\rho_{\text{init}}\rangle= \prod_{i=1}^{L-1}\left\vert \tau_{i+1/2}^{z}=+1\right\rangle=\left|s=\frac{L-1}{2},m=\frac{L-1}{2}\right\rangle$.

\textit{$\mathfrak{su}(2)$ algebra approach---}Within the $\mathfrak{su}(2)$ representation, the action of the Hamiltonian on the state $\left\vert s=\frac{L-1}{2},m=s-n\right\rangle$ is given by (recall that $S^{\pm}\vert s,m\rangle = \sqrt{(s\mp m)(s\pm m+1)}\vert s,m\pm1\rangle$)
\begin{align}
H^{\text{NN}}|s,s-n\rangle=&-\sqrt{n(L-n)}|s,s-n+1\rangle-\sqrt{(n+1)(L-n-1)}|s,s-n-1\rangle.
\label{eq:Hamilotnian for target 1}
\end{align}
By comparing this with the three-term recurrence relation, we obtain the correspondence: $\vert s=(L-1)/2,m=s-n\rangle\rightarrow (-1)^n \vert K_n\rangle$, and the Lanczos coefficients are read as $a_n=0$ and $b_n=\sqrt{n(L-n)}$, where $n=0,1,\cdots,L-1$. Notice that in this construction the initial state $\vert\rho_{\text{init}}\rangle$ must correspond to the highest-weight state, ensuring its mapping to $\vert K_{0}\rangle$.

\textit{Moment approach---}Given the Hamiltonian $H^{\text{NN}}$ and the initial state, either before or after the Kramers--Wannier transformation, the correlation function can be straightforwardly evaluated as $C(\tau)=\langle\rho_{\text{init}}\vert e^{-H^{\text{NN}}\tau} \vert\rho_{\text{init}}\rangle=\cosh^{L-1}{(\tau)}$, for which the moments are obtained as
\begin{align}
    \mu_{n}=\frac{d^{n}C(\tau)}{d\tau^{n}}\bigg|_{\tau=0}
    =\sum_{k=0}^{L-1} \frac{1}{2^{L-1}} {L-1\choose k} (2k-L+1)^{n}.
\end{align}
These moments can be generated by the Krawtchouk orthonormal polynomials~\cite{moment}, from which the Lanczos coefficients can be directly identified as
\begin{align}
    a_{n}=0;\quad
    b_{n}=\sqrt{n(L-n)},
\end{align}
where $n=0,1,\cdots,L-1$.

\section{Analytical Solution of Krylov Complexity for $H^{\text{NN}}$}
In order to calculate the Krylov complexity for $H^{\text{NN}}$, we begin by evaluating the imaginary time evolution of 
Eq.~(\ref{eq:SU(2) Hamiltonian}),
\begin{align}
|\rho(\tau)\rangle=e^{-H^{\text{NN}}\tau} \prod_{i=1}^{L-1}\left\vert \tau_{i+1/2}^{z}=+1\right\rangle=e^{(S^++S^-)\tau}|s,s\rangle,
\label{eq:tevo-SU(2)}
\end{align}
where $s=(L-1)/2$.
Using the Baker--Campbell--Hausdorff formula, the
time evolution operator can be decomposed as
\begin{align}
e^{(S^++S^-)\tau}=e^{a_-S^-}e^{a_0S^z}e^{a_+S^+},
=\label{eq:BCH}
\end{align}
where $a_+=-\tanh(\tau)$, $a_0=2\ln{[\cosh(\tau)]}$, and $a_-=\tanh(\tau)$.
Consequently, the unnormalized $|\rho(\tau)\rangle$ becomes
\begin{align}
|\rho(\tau)\rangle&=(\cosh(\tau))^{2s}e^{a_-S^-}|s,s\rangle=\sum_{n=0}^{L-1}\frac{\tanh^n(\tau)}{\cosh^{-(L-1)}(\tau)}\sqrt{{L-1\choose n}}|s,s-n\rangle=\sum_{n=0}^{L-1}\frac{(-1)^n\tanh^n(\tau)}{\cosh^{-(L-1)}(\tau)}\sqrt{{L-1\choose n}}|K_n\rangle,
\label{eq:tevo-SU(2)-2}
\end{align}
from which we identify the amplitude $\psi(\tau)$ and compute the normalization factor, 
\begin{align}
\sum_{n=0}^{n=L-1}|\psi_n|^2=\sum_{n=0}^{n=L-1} {L-1\choose n}\gamma^n(1+\gamma)^{L-1-n}
=(1+2\gamma)^{L-1},
\label{eq:denominator of tevo-SU(2)}
\end{align}
where $\gamma=\sinh^2(\tau)$. The Krylov complexity then follows as
\begin{align}
K(\tau)
=\frac{
\sum_{n=0}^{n=L-1} n{L-1\choose n}\gamma^n(1+\gamma)^{L-1-n}}{(1+2\gamma)^{L-1}}
=(L-1)\frac{\gamma}{1+2\gamma},
\label{eq:K-complexity SU(2)}
\end{align}
which reproduces Eq.~(\ref{eq:K_NN}) in the main text.

\section{Details of Numerical Calculation for Rényi-2 Correlator}
\begin{figure}[t]
\begin{center} 
\vspace{0.5cm}
\includegraphics[width=10.5cm]{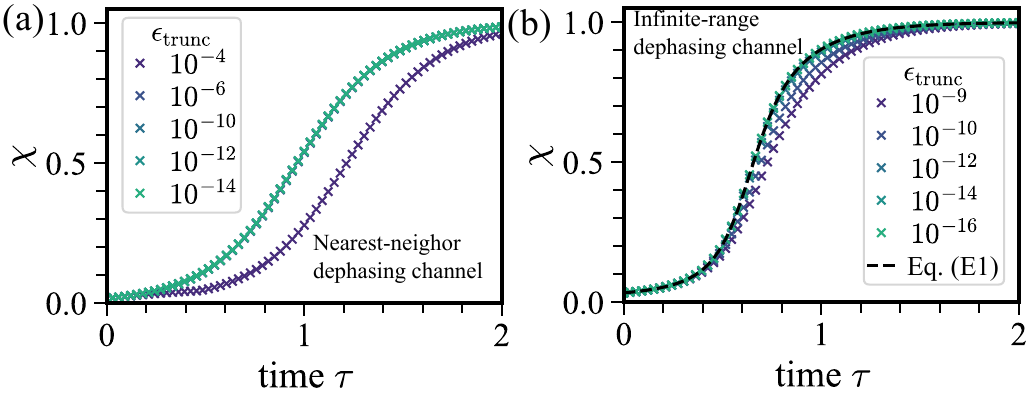}  
\end{center} 
\caption{The truncation cutoff $\epsilon_{\rm trunc}$ dependence of $\chi$. (a) $H^{\text{NN}}$ with $L=60$. (b) $H^{\text{IR}}$ with $L=30$. 
For either case, one can see that $\chi$ successfully converges when the truncation cut-off is set to a sufficiently small value, such as $10^{-14}$. In the case of $H^{\text{IR}}$, the results obtained using the tensor-network method for small truncation cut-off align well with the result obtained by Eq.~(E1).
}
\label{FigA1}
\end{figure}

In this section, we provide details of the numerical calculations for the Rényi-2 correlator. We first prepare the initial state in the matrix product state representation combined with the Choi map.
The imaginary time evolution is then performed according to Eqs.~(\ref{eq:nn-decohere-double}) and (\ref{eq:IR-decohere-double}).
Since all terms in $H^{\text{NN}}$ and $H^{\text{IR}}$ mutually commute, each term can be evolved independently without loss of accuracy.
In our simulation, choosing a sufficiently small truncation cutoff ensures the convergence of the Rényi-2 correlator successfully.
In addition, for $H^{\text{IR}}$,  the Rényi-2 correlator can be computed directly using the Krylov subspace method,
\begin{align}
\chi=&\frac{\sum_{ij}\langle \rho(\tau)|Z_i^uZ_i^\ell Z_{j}^uZ_{j}^\ell|\rho(\tau)\rangle}{L^2 \langle \rho(\tau)|\rho(\tau)\rangle}\nonumber\\
=&-\frac{\sum_{n,n'}\langle K_{n'}|2H^{\text{IR}}-L|K_n\rangle\psi_{n'}(\tau)\psi_n(\tau)}{L\sum_n|\psi_n(\tau)|^2}\nonumber\\
=&-\frac{\sum_{n}[b_{n+1}\psi_{n+1}(\tau)\psi_n(\tau)+(a_n-L)\psi_{n}(\tau)\psi_n(\tau)+b_{n-1}\psi_{n-1}(\tau)\psi_n(\tau)]}{L\sum_n|\psi_n(\tau)|^2}.
\end{align}

Figure~\ref{FigA1} shows the Rényi-2 correlator obtained with different truncation cutoffs. We find that setting the cutoff to values around $10^{-14}$ is sufficient to achieve convergence.

\section{Definition of Infinite-Range Dephasing Channel}
In this work, we use a modified definition of the infinite-range dephasing channel studied in Ref.~[55]:
 \begin{equation}
 \mathcal{E}_{ij}[\rho]=\frac{1}{2}\left(1+e^{-\frac{2\tau}{L}}\right)\rho+\frac{1}{2}(1-e^{-\frac{2\tau}{L}})Z_iZ_j\rho Z_iZ_j,
 \label{eq:IR}
 \end{equation}
where $\tau$ lies in the range of $0\le\tau<\infty$, and $\frac{1}{2}(1\pm e^{-\frac{2\tau}{L}})\in[0,\frac{1}{2})$.
Using the doubled Hilbert space formalism (briefly reviewed in SM.~\ref{sm:decoherence_channel_to_imaginary_t_evolution}) and the identity Eq.~(\ref{eq:exp_expansion_identity}),
Eq.~(\ref{eq:IR}) can be recast to
 \begin{align}
 \mathcal{E}|\rho\rangle=&
 \prod_{i<j}\left[\frac{1}{2}\left(1+e^{-\frac{2\tau}{L}}\right) I_i^uI_{i+1}^uI_{i}^\ell I_{i+1}^\ell+\frac{1}{2}\left(1-e^{-\frac{2\tau}{L}}\right)Z_i^uZ_{j}^u Z_i^\ell Z_{j}^\ell\right]|\rho\rangle  \nonumber \\
 =&e^{- H^{\text{IR}} \tau}|\rho\rangle,
 \end{align}
 where $H^{\text{IR}}=-\sum_{i<j} \left(Z_i^u Z_i^l Z_j^u Z_j^l -1 \right)/L$ is the infinite-range Ising Hamiltonian. We remark that this mapping is exact.

\section{Lanczos Coefficients of $H^{\text{IR}}$}
We here provide the detailed derivation of the Lanczos coefficients for the Lipkin--Meshkov--Glick Hamiltonian $H^{\text{IR}}=-\frac{2}{L}S_{x}^{2}+\frac{L}{2}$, with the initial state $\prod_{i}\vert\Uparrow_{i}\rangle$. Expanding $H^{\text{IR}}$ in the spin basis $\vert s,m\rangle$ using $S^{\pm}\vert s,m\rangle = \sqrt{(s\mp m)(s\pm m+1)}\vert s,m\pm1\rangle$, we obtain (recall that $s=L/2$)
\begin{align}
    H^{\text{IR}}\vert s,m\rangle =& C^{+2}(s,m)\vert s,m+2\rangle
    +C^{0}(s,m)\vert s,m\rangle
    +C^{-2}(s,m)\vert s,m-2\rangle,
 \label{eq:App_F1}
\end{align}
where
\begin{align}
    C^{+2}(s,m) =& -\frac{1}{2L}\sqrt{(s-m)(s+m+1)(s-m-1)(s+m+2)},\\
    C^{0}(s,m) =& -\frac{1}{2L}(2s^{2}-2m^{2}+2s)+\frac{L}{2},\\
    C^{-2}(s,m) = & -\frac{1}{2L}\sqrt{(s+m)(s-m+1)(s+m-1)(s-m+2)}.
\end{align} 
To connect the spin basis with the Krylov basis, we first note that the initial state---the highest weight state $\vert L/2,L/2\rangle$---corresponds to the zeroth Krylov vector $\vert K_{0}\rangle$. Since the basis $\vert s,m\rangle$ can change in $m$ only by $\pm2$, due to the quadratic form of $H^{\text{IR}}$, the subsequent Krylov states are naturally identified as $\vert L/2,L/2-2\rangle\mapsto-\vert K_{1}\rangle$, $\vert L/2,L/2-4\rangle\mapsto\vert K_{2}\rangle$, and so on. In general,
\begin{align}
    \left\vert \frac{L}{2},\frac{L}{2}-2n\right\rangle \mapsto (-1)^n\vert K_{n}\rangle,
\end{align}
where $n=0,1,\cdots,L/2$. Note that the factor $(-1)^{n}$ is introduced to ensure that $b_{n}$ follows the positive convention. By setting $s=L/2$ and $m=L/2-2n$ and using this map, Eq.~(\ref{eq:App_F1}) becomes
\begin{align}
    H^{\text{IR}}\vert K_{n}\rangle = -C^{2}\left(\frac{L}{2},\frac{L}{2}-2n\right) \vert K_{n-1}\rangle
    + C^{0}\left(\frac{L}{2},\frac{L}{2}-2n\right) \vert K_{n}\rangle
    - C^{-2}\left(\frac{L}{2},\frac{L}{2}-2n\right) \vert K_{n+1}\rangle\label{eq:H_IR-tridiagonal},
\end{align}
which takes the standard three term recurrence form. Accordingly, by comparing with Eq.~(\ref{eq:recurrence_eq}) in the main text, we straightforwardly read $a_{n}=C^{0}(L/2,L/2-2n)$ and $b_{n}=-C^{2}(L/2,L/2-2n)$, which are Eqs.~(\ref{eq:H_IR_an}) and (\ref{eq:H_IR_bn}) therein.

\section{Holstein--Primakoff Transformation and the Area Law Regime of $\mathcal{K}(\tau)$}
We present the detailed derivation of the Krylov complexity in the thermodynamic limit, where $\mathcal{K}(\tau)$ exhibits an area law scaling for $0<\tau<1/2$, corresponding to Eq.~(\ref{eq:area_law_K}) in the main text. Our objective is to evaluate $\psi_n(\tau)=\langle K_{n}\vert\rho(\tau)\rangle/(\langle\rho(\tau)\vert\rho(\tau)\rangle)^{1/2}$, where $\vert\rho(\tau)\rangle=e^{-H^{\text{IR}}\tau}\vert K_{0}\rangle$.

To this end, we employ the standard Holstein--Primakoff transformation, expressing the spin operators in terms of bosonic creation and annihilation operators,
\begin{align}
    S^{+}\mapsto&\sqrt{2s}\sqrt{1-\frac{a^{\dagger}a}{2s}}a,\\
    S^{-}\mapsto&\sqrt{2s}\sqrt{1-\frac{a^{\dagger}a}{2s}}a^{\dagger},\\
    S^{z}\mapsto&s-a^{\dagger}a,\\
    \vert s,s-\lambda\rangle\mapsto&\frac{1}{\sqrt{n!}}\left(a^{\dagger}\right)^{\lambda}\vert0\rangle \equiv
    \vert\lambda\rangle
\end{align}
As a result, the relevant state becomes $\vert K_{n}\rangle = (-1)^n\vert L/2,L/2-2n\rangle \mapsto (-1)^n\left(a^{\dagger}\right)^{2n}\vert0\rangle/\sqrt{(2n)!}$. In the thermodynamic limit---equivalently the large spin limit ($s=L/2$)---we expand $H^{\text{IR}}$ to the leading order of $s$,
\begin{align}
    H^{\text{IR}} - \frac{L}{2}=&-\frac{2S_{x}^{2}}{L}\notag\\
    =&-\frac{1}{2L}\left(S^{+}S^{+}+S^{+}S^{-}+S^{-}S^{+}+S^{-}S^{-}\right)\notag\\
    \simeq& -\frac{1}{2L}2s (a+a^{\dagger})^{2} = -\frac{1}{2}(a+a^{\dagger})^{2},\label{AppendHIR}
\end{align}
where $s$ and $L$ cancel in the final step, leading to an the $L$-independent Hamiltonian. Consequently, the numerator of $\psi_n(\tau)$ becomes
\begin{align}
    \langle K_{n}\vert\rho(\tau)\rangle =& \langle K_{n}\vert e^{-H^{\text{IR}}\tau}\vert K_{0}\rangle\notag\\
    \simeq&(-1)^n e^{-\frac{L}{2}\tau} \langle 2n\vert e^{\frac{1}{2}(a+a^{\dagger})^{2}\tau}\vert 0\rangle\notag\\
    =& (-1)^ne^{-\frac{L}{2}\tau}\langle 2n\vert e^{x^{2}\tau}\vert0\rangle\notag\\
    =& \frac{(-1)^ne^{-\frac{L}{2}\tau}}{\sqrt{2^{2n}(2n)!\pi}} \int dx\,e^{(\tau-1)x^{2}}H_{2n}(x),
\end{align}
where we have introduced the standard harmonic position operator $x=(a+a^{\dagger})/\sqrt{2}$, and $H_{\nu}(x)$ denotes the physicists' Hermite polynomial. The integral admits an exact solution when $0<\tau<1$,
\begin{align}
    \langle K_{n}\vert\rho(\tau)\rangle \simeq
    e^{-\frac{L}{2}\tau} \frac{\sqrt{(2n)!}}{2^{n}n!} \sqrt{\frac{1}{1-\tau}} \left(\frac{-\tau}{1-\tau}\right)^{n}.
\end{align}
Similarly, the denominator becomes
\begin{align}
    \left(\langle\rho(\tau)\vert\rho(\tau)\rangle\right)^{\frac{1}{2}} \simeq
    e^{-\frac{L}{2}\tau} \left(\frac{1}{1-2\tau}\right)^{\frac{1}{4}}\label{eq:norm}
\end{align}
with a narrower convergence domain $0<\tau<1/2$. Thus, the amplitude becomes
\begin{align}
    \psi_n(\tau) \simeq \frac{\sqrt{(2n)!}}{2^{n}n!}\sqrt{\frac{1}{1-\tau}}\left(\frac{-\tau}{1-\tau}\right)^{n}\left(\frac{1}{1-2\tau}\right)^{-1/4}.\label{eq:H9}
\end{align}
Using Eq.~(\ref{eq:K-complexity}) in the main body of the text, the corresponding Krylov complexity is
\begin{align*}
    \mathcal{K(\tau)}=&\sum_{n=0}^{\infty}n\abs{\psi_n(\tau)}^{2}
    =\sum_{n=0}^{\infty}n{2n \choose n}\left(\frac{\eta(\tau)}{4}\right)^{n},
\end{align*}
where $\eta(\tau) \equiv [\tau/(1-\tau)]^{2}$. The infinite summation can be evaluated by using a generating function,
\begin{align}
    &\sum_{n=0}^{\infty}{2n \choose n}x^{n} = \frac{1}{\sqrt{1-4x}}\notag\\
    \Rightarrow&\sum_{n=0}^{\infty}n{2n \choose n}x^{n} = 2x(1-4x)^{-\frac{3}{2}},
\end{align}
which converges for $x<1/4$. Thus,
\begin{align}
    \mathcal{K}(\tau) = \frac{\eta(\tau)}{2} (1-\eta(\tau))^{-\frac{3}{2}}
    = \frac{\tau^{2}}{2(1-2\tau)}\label{eq:H11},
\end{align}
with the convergence condition $\tau<1/2$. Using the Holstein--Primakoff transformation, we recover the area law behavior of $\mathcal{K}(\tau)$, which is valid within $0<\tau<1/2$, and undergoes a divergence at $\tau=1/2$, signaling a phase transition.

\section{The Volume Law Regime of $\mathcal{K}(\tau)$}
To explore the Krylov complexity $\mathcal{K}(\tau)$ beyond the range $0<\tau<1/2$, we follow a different pathway. Notice that $S_{x}$ in $H^{\text{IR}}$ is not diagonal in the spin basis $\vert s,m\rangle$, but becomes diagonal in the rotated spin basis $\vert s,m\rangle_{x}$ which satisfies $S_{x}\vert s,m\rangle_{x} = m\vert s,m\rangle_{x}$. The two bases are related by a spin rotation, $\vert s,m\rangle_{x} = e^{-i\pi S_{y}/2}\vert s,m\rangle$. As a result,
\begin{align}
    \vert s,m\rangle =& \sum_{m'=-s}^{s}\vert s,m'\rangle_{x}\,{}_{x}\langle s,m'\vert s,m\rangle\notag\\
    =& \sum_{m'=-s}^{s} d_{m'm}^{s}\left(-\frac{\pi}{2}\right) \vert s,m'\rangle_{x},
\end{align}
where $d_{m'm}^{s}(\theta) \equiv \langle s,m'\vert e^{-i\theta S_{y}}\vert s,m\rangle$ is the so-called Wigner small d-matrix~\cite{sakurai2020modern}.

To evaluate $\psi_n(\tau) = \langle K_{n}\vert \rho(\tau)\rangle/(\langle\rho(\tau)\vert\rho(\tau)\rangle)^{1/2}$ where $\vert\rho(\tau)\rangle = e^{-H^{\text{IR}}\tau}\vert K_{0}\rangle$, we begin by focusing on the matrix element,
\begin{align}
    &\langle s,m\vert e^{-H^{\text{IR}}\tau} \vert s,n\rangle\notag\\
    =& e^{-\frac{L}{2}\tau} \sum_{m'=-s}^{s} d_{m'm}^{s}\left(-\frac{\pi}{2}\right) d_{m'n}^{s}\left(-\frac{\pi}{2}\right) e^{\frac{2m'^{2}}{L}\tau}.\label{eq:matrix_element_in_terms_of_Wigner_d}
\end{align}
We set $s=n=L/2$ and $m=L/2-2n$ for the numerator $\langle K_{n}\vert\rho(\tau)\rangle = (-1)^n\langle L/2,L/2-2n\vert e^{-H^{\text{IR}}\tau}\vert L/2,L/2\rangle$, and we take $s=m=n=L/2$ and replace $\tau\rightarrow2\tau$ for denominator $(\langle\rho(\tau)\vert\rho(\tau)\rangle)^{1/2} = \left( \langle L/2,L/2\vert e^{-2H^{\text{IR}}\tau}\vert L/2,L/2\rangle \right)^{1/2}$. The Wigner d-matrix has a general closed form~\cite{sakurai2020modern},
\begin{align}
     d_{m'n}^{s}(\theta) = \sum_{k=k_{\text{min}}}^{k=k_{\text{max}}} (-1)^{k-n+m'} \frac{\sqrt{(s+n)!(s-n)!(s+m')!(s-m')!}}{(s+n-k)!(s-k-m')!(k-n+m')!k!} \left(\cos{\frac{\theta}{2}}\right)^{2s-2k+n-m'} \left(\sin{\frac{\theta}{2}}\right)^{2k-n+m'},
\label{eq:d-matrix}
\end{align}
where the summation is over all integer $k$ values for which the factorial arguments in the denominator are nonnegative. While Eq.~(\ref{eq:d-matrix}) appears cumbersome, it simplifies considerably in specific cases. In particular, when $n=s$ (the case of interest here), the Wigner d-matrix reduces to
\begin{align}
    d_{m's}^{s} = \sqrt{{2s \choose s+m'}} \left(\cos{\frac{\theta}{2}}\right)^{s+m'} \left(\sin{\frac{\theta}{2}}\right)^{s-m'}.\label{eq:Wigner_d_special_case}
\end{align}
Using Eqs.~(\ref{eq:matrix_element_in_terms_of_Wigner_d}) and (\ref{eq:Wigner_d_special_case}), we obtain the amplitude,
\begin{align}
    \psi_n(\tau) = 
    \frac{(-1)^n\sum_{m'=-\frac{L}{2}}^{\frac{L}{2}} d_{m',\frac{L}{2}-2n}^{\frac{L}{2}}\left(\frac{\pi}{2}\right) \sqrt{\displaystyle\binom{L}{\frac{L}{2}+m'}} e^{\frac{2m'^{2}}{L}\tau}}
    {\sqrt{\sum_{m'=-\frac{L}{2}}^{\frac{L}{2}} {L \choose \frac{L}{2}+m'} e^{\frac{4m'^{2}}{L}\tau}}},
\label{eq:H5}
\end{align}
where we have utilized properties $d_{m'n}^{s}(-\theta) = d_{nm'}^{s}(\theta)=(-1)^{m'-n}d_{m'n}^{s}(\theta)$. Note that Eq.~(\ref{eq:H5}) is exact.

To investigate the volume law regime, we consider the limit of large $L$ and large $\tau$ while keeping the ratio $\tau/L$ fixed. In this limit, the summations in Eq.~(\ref{eq:H5}) are dominated by the contributions from the high spin terms $m'=\pm L/2$. Therefore, the amplitude $\psi_n(\tau)$ simplifies to
\begin{align}
    \psi_n(\tau) \simeq (-1)^n\sqrt{{L \choose 2n} 2^{1-L}}\label{eq:I6},
\end{align}
where we have utilized
\begin{align}
    d^{\frac{L}{2}}_{\frac{L}{2},\frac{L}{2}-2n} \left(\frac{\pi}{2}\right) =
    \sqrt{{L\choose 2n}} 2^{-\frac{L}{2}}.
\end{align}

Using Eq.~(\ref{eq:K-complexity}) in the main body of the text, we then compute the corresponding Krylov complexity,
\begin{align}
    \mathcal{K}=\sum_{n}n\abs{\psi_n(\tau)}^{2}
    =\sum_{n=0}^{L/2} n {L \choose 2n} 2^{1-L}
    =\sum_{\nu=0,2,\cdots,L}\frac{\nu}{2} {L \choose \nu} 2^{1-L}.
\end{align}
This sum can be analytically solved by using a generating function,
\begin{align}
    &\sum_{\nu=0}^{L} {L \choose \nu} x^{\nu} = (1+x)^{L}\notag\\
    \Rightarrow&\sum_{\nu=0}^{L} \nu {L \choose \nu} x^{\nu-1} = L(1+x)^{L-1} \equiv D(x)\notag\\
    \Rightarrow& \frac{D(1)-D(-1)}{2} = \sum_{\nu=0,2,\cdots,L}\nu{L \choose \nu} = \frac{L 2^{L-1}}{2}\notag\\
    \Rightarrow&\mathcal{K} = \frac{L}{4}\label{eq:K_in_volume},
\end{align}
which precisely matches our numerical results.

\section{The Relation between Krylov Complexity and Expectation Value}
In this section, we clarify the relation between Krylov complexity and the expectation value $\langle S_x^2 \rangle$, which is relevant to the infinite-range dephasing channels. In particular, we derive the area-law result, Eq. (\ref{eq:connection_K_Sx2_area}), and volume-law result, Eq. (\ref{eq:connection_K_Sx2_volumn}), quoted in the End Matter.

To relate $\langle S_x^2\rangle$ to the Krylov complexity $\mathcal{K}$, note that $H^{\text{IR}}=2S_x^2/L -L/2$, which implies
\begin{align}
    \left\langle \frac{S_x^2}{L^2} \right\rangle = \frac{1}{2L}\langle H^{\text{IR}}\rangle + \frac{1}{4}
    = -\frac{1}{2}\frac{d}{d\tau}\ln{\langle\rho(\tau)\vert\rho(\tau)\rangle}
    = \frac{1}{2L} \frac{\langle K_0\vert H^{\text{IR}} e^{-H^{\text{IR}}2\tau} \vert K_0\rangle}{\langle\rho(\tau\vert\rho(\tau)\rangle} + \frac{1}{4}.\label{eq:relation_bt_Sx2_and_HIR}
\end{align}

In the area-law regime, we may apply the Holstein--Primakoff transformation in the thermodynamic limit. Using Eq.~(\ref{eq:norm}), we obtain
\begin{align}
    \langle H^{\text{IR}}\rangle = -\frac{1}{2}\frac{d}{d\tau}\ln[\langle\rho(\tau)|\rho(\tau)\rangle]
    \simeq\frac{L}{2}-\frac{1}{2(1-2\tau)},\label{eq:expt_HIR}
\end{align}
which then gives
\begin{align}
    \left\langle \frac{S_x^2}{L^{2}}\right\rangle =\frac{1}{4(1-2\tau)L}.\label{eq:expt_SX2}
\end{align}
Comparing with Eq.~(\ref{eq:H11}), we recover Eq. (\ref{eq:connection_K_Sx2_area}) in the End Matter.

In the volume-law regime, the numerator on the very RHS of Eq.~(\ref{eq:relation_bt_Sx2_and_HIR}) can be obtained from Eq.~               (\ref{eq:matrix_element_in_terms_of_Wigner_d}) as a generating function: setting $m=n=0$, differentiating with respect to $\tau$, (and then replacing $\tau$ by $2\tau$ later), we find
\begin{align}
    &\langle K_0\vert H^{\text{IR}} e^{-H^{\text{IR}}\tau} \vert K_0\rangle =
    -\frac{d}{d\tau} \langle K_0\vert e^{-H^{\text{IR}}\tau} \vert K_0\rangle\notag\\
    =& \frac{L}{2} e^{-L\tau/2} \sum_{m'=-\frac{L}{2}}^{\frac{L}{2}} {L \choose \frac{L}{2}+m'} 2^{-L} e^{2m'^2\tau/L}
    -e^{-L\tau/2} \sum_{m'=-\frac{L}{2}}^{\frac{L}{2}} {L \choose \frac{L}{2}+m'} 2^{-L} \frac{2m'^2}{L} e^{2m'^2\tau/L},
\end{align}
In the large-$L$ and large-$\tau$ limit, the summations are dominated by $m'=\pm L/2$, leading to
\begin{align}
    \langle K_0\vert H^{\text{IR}} e^{-H^{\text{IR}}\tau} \vert K_0\rangle \simeq
    2 \left( \frac{L}{2} e^{-L\tau/2} 2^{-L} e^{L\tau/2} - e^{-L\tau/2} 2^{-L} \frac{L}{2} e^{L\tau/2} \right) = 0.
\end{align}
Thus,
\begin{align}
    \left\langle \frac{S_x^2}{L^2} \right\rangle = \frac{1}{4},
\end{align}
which, after comparing to Eq. (\ref{eq:K_complexity_volume_law}) in the main text, recovers Eq. (\ref{eq:connection_K_Sx2_volumn}) in the End Matter.

\end{document}